\documentclass[11pt,twocolumn]{IEEEtran}
\usepackage{amssymb, amsmath, graphicx, epsfig, epstopdf,algorithm,algorithmic,amsthm,color}
\usepackage[mathscr]{euscript}
\usepackage[nospace]{cite}

\usepackage{epsfig,amsmath,amssymb}%,hyperref}
\usepackage{enumerate}
\usepackage{subfigure}
\usepackage{graphicx} 
\usepackage{epstopdf}
\usepackage{xcolor}

\usepackage{amsmath,amssymb}
\usepackage{subfigure,algorithm,algorithmic}

\catcode`\@=11
\begingroup
\catcode`|=0\catcode`[=1
\catcode`]=2\catcode`\{=12\catcode`\}=12
\catcode`\\=12
|long|gdef|@sxcomment#1\end{comment}[|end[comment]]
|endgroup
\def\comments{
\def\@comment{\bgroup\let\do\@makeother\dospecials}
\def\comment{\@comment\@sxcomment}
\let\endcomment=\egroup}
\comments
\catcode`\@=12

\newcommand{\define}{\triangleq}
\def\x{{\mathbf{ x}}}
\usepackage{amsmath}

\newcommand{\y}{\mathbf{y}}

\newcommand{\e}{\mathbf{e}}

\newcommand{\R}{\mathbf{R}}
\newcommand{\ba}{\mathbf{a}}
\newcommand{\bb}{\mathbf{b}}
\newcommand{\A}{\mathbf{A}}
\newcommand{\bQ}{\mathbf{Q}}

\newcommand{\W}{\mathbf{W}}

\newcommand{\z}{\mathbf{z}}
\newcommand{\br}{\mathbf{r}}
\newcommand{\bu}{\mathbf{u}}

\newcommand{\bp}{\mathbf{p}}
\newcommand{\tbp}{\tilde{\bp}}

\newcommand{\bP}{\mathbf{P}}
\newcommand{\bB}{\mathbf{B}}
\newcommand{\bT}{\mathbf{T}}

\newcommand{\bsigma}{\boldsymbol{\sigma}}

        \makeatletter
        \def\fps@eqnfloat{!t}
        \def\ftype@eqnfloat{4}
        
        \newenvironment{eqnfloat*}
               {\@dblfloat{eqnfloat}}
               {\end@dblfloat}
        \makeatother
\makeatletter
\let\oldabs\abs
\def\abs{\@ifstar{\oldabs}{\oldabs*}}
\let\oldnorm\norm
\def\norm{\@ifstar{\oldnorm}{\oldnorm*}}
\makeatother

\renewenvironment{thebibliography}[1]{%
 \begin{oldthebibliography}{#1}%
 \setlength{\parskip}{0ex}%
 \setlength{\itemsep}{0ex}%
}%
{%
\end{oldthebibliography}%
}%
\newtheorem{theorem}{Lemma}

\title{Generalized Sparse Covariance-based Estimation\thanks{This work was supported in part by the Swedish Research Council and the Crafoord's and Carl Trygger's foundations. This work has been presented in part at the ICASSP conference 2017.}}
\author{Johan Sw\"ard,~\IEEEmembership{Student member,~IEEE},\thanks{The authors are with the Department of Mathematical Statistics, Lund University, 221 00 Lund, Sweden (emails: \texttt{\{js, sia, aj\}@maths.lth.se)}.} Stefan Ingi Adalbj\"ornsson,~\IEEEmembership{Member,~IEEE}, and 
Andreas~Jakobsson,~\IEEEmembership{Senior Member,~IEEE}}
\begin{document}

\maketitle

\begin{abstract}
In this work, we extend the sparse iterative covariance-based estimator (SPICE), by generalizing the formulation to allow for different norm constraints on the signal and noise parameters in the covariance model. For a given norm, the resulting extended SPICE method enjoys the same benefits as the regular SPICE method, including being hyper-parameter free, although the choice of norms are shown to govern the sparsity in the resulting solution. Furthermore, we show that solving the extended SPICE method is equivalent to solving a penalized regression problem, which provides an alternative interpretation of the proposed method and a deeper insight on the differences in sparsity between the extended and the original SPICE formulation. We examine the performance of the method for different choices of norms, and compare the results to the original SPICE method, showing the benefits of using the extended formulation. We also provide two ways of solving the extended SPICE method; one grid-based method, for which an efficient implementation is given, and a gridless method for the sinusoidal case, which results in a semi-definite programming problem.
\end{abstract}
%
%\begin{keyword}
%Covariance fitting, sparse reconstruction, convex optimization
%\end{keyword}
%
%\end{frontmatter}

\section{Introduction}
%\noindent {\em Notation:} In this paper, vectors (matrices) will be denoted by lower (upper) case boldface letters. The diag$(\x)$ operator takes the vector $\x$ and inserts it on the diagonal of a matrix with zeros elsewhere. Furthermore, $(\cdot)^T$ and $(\cdot)^*$ denotes the transpose and the complex conjugate transpose respectively, and tr$(\mathbf{X})$ denotes the trace of the matrix $\mathbf{X}$. We abbreviate with respect to as w.r.t and use SNR for short for signal to noise ratio.
%\\ \newline
\noindent
Many problems in signal processing may be well described using a linear model, such that
\begin{align}\label{eq:model}
\y=\bB\x+\e 
\end{align}
where $\y\in \mathbb{C}^N$ is a vector of measurements, $\bB$ a matrix of regressors, $\x$ the parameter vector, and $\e$ denotes an additive (complex-valued) noise term, typically assumed to have zero mean and covariance matrix $\mathbf{\Sigma}$. This model occurs in a wide range of applications, such as in, e.g., audio and speech processing \cite{ChristensenJ09, AdalbjornssonJC15_109} and spectroscopy \cite{LiRL98_45, SunS12_60,SomasundaramJSA07,TanTC04_11,SwardAJ16_128}.

Historically, there have been two main principles available for solving these kinds of problems: {\em parametric} and {\em non-parametric} methods. The latter approach does not rely on any {\textit{a-priori}} information about the signal, including assumptions on the model structure or order, and such techniques are therefore more robust to uncertainties in such model assumptions than the former. However, this robustness comes with the downside that the non-parametric methods are, in general, not able to yield as good performance as the parametric approaches, which typically in turn are less robust \cite{StoicaM05}. Recently, notable efforts have been made to combine these two approaches, developing so-called {\em semi-parametric} approaches, which typically only make some weak model structure assumptions, such that assuming that the solution is sparse, although restrain from making any stronger model order assumptions.
This implies that although the dictionary, $\bB\in \mathbb{C}^{N\times M}$, is formed using $M\gg N$ signal candidates, only a few of these candidates are assumed present in the signal. The problem is thus transformed into finding the subset of these $M$ candidates best approximating the measured signal $\y$. Many sparse methods do this by enforcing sparsity on the vector $\x$, creating a trade-off between the model fit and the level of sparsity. In \cite{Tibshirani96_58}, this was done by introducing the LASSO optimization problem
\begin{align}\label{eq:LASSO}
\underset{\x}{\text{minimize}}\ \frac{1}{2}\left|\left|\y-\bB\x \right|\right|_2^2+\mu||\x||_1
\end{align}
where the first term penalizes the {$\ell_2$-norm} distance between the model and the signal, and the second term enforces sparsity upon the vector $\x$, with $\mu$ being a user parameter that governs the trade-off between the two terms. Recently, many other sparse methods have been proposed (see, e.g.,\cite{CandesWB08_14,TibshiraniSRZK05_67,FangLSLL14_21,GorodnitskyR97_45,TanRLS11_59,HastieTW15} and the references therein). One potential drawback of these methods is the requirement of selecting the user parameter, which is often a non-trivial task.
Sometimes there are physical aspects that may aid in the choice of this parameter, whereas, in other, some kind of rule of thumb on how to choose it may be found \cite{SwardBJH16_64}. Other ideas include solving the problem for all different values of the parameter \cite{EfronHJT04_32,HastieTW15}, or to use some iterative process for aiding in the choice \cite{CandesWB08_14,WipfN10_4,DaubechiesDFG10_63}. Another common way is to use cross-validation to find a suitable regularization parameter (see, e.g., \cite{HastieTW15}).

In \cite{StoicaBL11_59}, a novel approach to form a sparse solution was proposed based on a covariance fitting criteria, and was shown to overcome the drawback of selecting the user parameter (see also \cite{StoicaBJ11_59b,StoicaB12_92,RojasKH13_61,StoicaZL14_33,ZachariahS15_63}). 
The minimization criteria that was proposed was
\begin{align}\label{eq:covFitSPICE}
\underset{{\tbp\geq 0}}{\text{minimize}}\ \left|\left|\R^{1/2}(\tbp)\left(\R(\tbp)-\y\y^*\right) \right|\right|^2_F
\end{align}
where $||\cdot||_F$ denotes the Frobenius norm, $(\cdot)^*$ the conjugate transpose, and where
\begin{align}\label{eq:R}
\R({\tbp}) &= \A\bP\A^*\\
\A &=\left[\begin{array}{c c c} \bB & \mathbf{I}\end{array}
\right] \\ \label{eq:pDef}
{\bp} &=  \left[\begin{array}{c c c c c c} p_1 & \dots & p_M \end{array} \right]^T\\\label{eq:sigmaDef}
\boldsymbol{\sigma} &= \left[\begin{array}{c c c} \sigma_1& \dots & \sigma_N
\end{array} \right]^T\\
\tbp &= \left[\begin{array}{c c c} \bp^T & \boldsymbol{\sigma}^T
\end{array} \right]^T\\
\bP &= \text{diag}\left( \tbp\right)
%\begin{bmatrix}
%p_1& & & & &0 \\
%& \ddots& &\\
%& & p_M &\\
%& & & \sigma_1\\
%& & & & \ddots &\\
%0& & & & &\sigma_N
%\end{bmatrix}
\end{align}
with $\mathbf{I}$ denoting the $N\times N$ identity matrix, $(\cdot)^T$ the transpose, { $\sigma_k$ the noise standard deviation} for sample $k$, and $\text{diag}(\z)$ the diagonal matrix with the vector $\z$ along its diagonal, and zeros elsewhere.
It was further shown that solving \eqref{eq:covFitSPICE} is equivalent with solving \cite{StoicaBL11_59}
\begin{align}\label{eq:SPICE}
\underset{ \tbp \geq 0}{\text{minimize}}\ \y^*\R^{-1}(\tbp)\y+||\tilde{\W}\tbp||_1
\end{align}
where 
\begin{align}
\tilde{\W} &= \text{diag}\left(\left[\begin{array}{c c c}
w_1 & \dots & w_{M+N}\end{array}\right]\right)\\
w_k &= ||\ba_k||_2^2/||\y||_2^2, \ \text{for } k=1,\dots,N+M 
\end{align}
with $\ba_k$ denoting the $k$th column of $\A$. By comparing \eqref{eq:LASSO} and \eqref{eq:SPICE}, it is clear that both problems minimize a signal fitting criteria, where the former more explicitly minimizes the distance between the model and the data, whereas the latter measures the distance through the inverse of the (model) covariance matrix. Furthermore, both problems include the $\ell_1$ norm, with the first one penalizing the parameters corresponding to the different candidates in the dictionary $\bB$, whereas the second, the so-called SPICE formulation, penalizes both the parameters corresponding to $\bB$ and the parameters corresponding to the noise.

In this paper, we generalize the SPICE approach to allow for different penalties on $\bp$ and $\boldsymbol{\sigma}$, as given in \eqref{eq:pDef} and  \eqref{eq:sigmaDef}, respectively, for two different cases; the first being when all $\sigma_k$ are equal, and the second when all $\sigma_k$ are allowed to differ. In the first case, we show that the choice of norm for the noise parameters corresponds to  different choices of the regularizing parameter, $\mu$, and the regularization norm, for a generalized form of the (weighted) square-root LASSO. In the case when all $\sigma_k$ are allowed to be different, the choices of norms are similarly shown to affect the sparsity level. This results in the fact that even if the different SPICE formulations are hyper-parameter free, one may interpret the choices of norms as the equivalence of selecting hyper-parameters dictating the sparseness of the solution, and that the original SPICE version is equivalent to one particular choice of norms.
We also provide an efficient grid-based implementation of the proposed method, which, indirectly, allows for solving (weighted) square-root LASSO problems for a wide choice of regularizing parameters. Additionally, we state a semi-positive programming (SDP) problem that allows for solving the proposed SPICE extension, for the sinusoidal case, without the use of a grid search.

\section{The $\{r,q\}$-SPICE formulation}\label{sec:II}
\noindent
It is worth noting that the second term in \eqref{eq:SPICE} penalizes the magnitude of each $p_j$ and $\sigma_k$, thus promoting a sparse solution with only a few of the terms in $ \tbp$ being non-zero. However, since the penalty does not distinguish between setting the different terms to zero, one may expect that some of the $\sigma_k$ may be forced to be zero as a part of the minimization. If one is interested in finding a sparse solution from the columns of the dictionary $\bB$ (in the same sense as in \eqref{eq:LASSO}), setting some of the noise parameters $\sigma_k$ to zero makes little sense. 
%
%Another implication from penalizing $\sigma_k$ in the same manner as $p_k$ arises from the fact that if $\R$ has not full rank then the inverse, as expressed in \eqref{eq:SPICE}, does not exist. 
Another intuition is given if one interprets \eqref{eq:SPICE} to require that $\R$ should be invertible. Assuming this, setting $\sigma_k$ to zero is problematic as the resulting covariance matrix, $\R$, loses rank, unless some of the $p_j$ are non-zero. Similar conclusions were stated in \cite{YangX15_63}, where a gridless formulation of SPICE where derived. It was shown that for the gridless version of SPICE, $\R$ had full rank with probability one, which in turn made the method overestimate the model order. Consequently, setting many $\sigma_k$ to zero will force the resulting $\bp$ to be less sparse, thus increasing the estimated model order. 
%To avoid these problems, one could instead make this implicit constraint explicit by reformulating the minimization as
%\begin{align}\label{eq:rank}
%\underset{\{p_k\}}{\text{minimize}}\ &\y^*\R^{-1}\y+||\tilde\W\tilde \bp||_1 \nonumber\\
%\text{subject to}\ &\text{rank}(\R)\geq N \nonumber\\
%&\R = \A\bP \A^*
%\end{align}
%Since $\A$ is chosen to have full rank, \eqref{eq:rank} implies that $\text{rank}(\bP)\geq N$. Thus, setting any  $\sigma_k$ to zero implies that at least one $p_k$ becomes non-zero. 
Thus, in the original SPICE formulation, $\sigma_k$ and $p_j$ are competing for the sparseness allowed in the solution of \eqref{eq:SPICE}. 

Alternatively, one could proceed to treat the $\sigma_k$ terms different from the rest of the $p_j$ terms. A naive way of doing this could be to omit $\sigma_k$ from the cost function of \eqref{eq:SPICE}, but this would result in all the $p_j$ terms being set to zeros as $\sigma_k$ may then take on any value which will make $\R$ full rank, and will thus make the $p_j$ terms redundant. Clearly, the $\sigma_k$ terms must instead be penalized to produce a meaningful solution to \eqref{eq:model}. This may be done in different ways, for instance using
\begin{align}\label{eq:newSPICE}
\underset{\bp\geq 0,\ { \bsigma \geq 0}}{\text{minimize}}\ \y^*\R^{-1}\y+||\W\bp||_r+||\W_\sigma\boldsymbol{\sigma} ||_q
\end{align}
where $r,q\geq1$, such that
\begin{align}
||\W\bp ||_r &= \left[ \sum^{M}_{k=1}w_{k}^r p_k^r\right]^{1/r}\\
||\W_\sigma\boldsymbol{\sigma} ||_q &= \left[ \sum^{N}_{k=1}w_{M+k}^q\sigma_k^q\right]^{1/q}\\
\W &= \text{diag}\left(\left[\begin{array}{c c c}
w_1 & \dots & w_{M}\end{array}\right]\right)\label{eq:W}\\
\W_\sigma &= \text{diag}\left(\left[\begin{array}{c c c}
w_{M+1} & \dots & w_{M+N}\end{array}\right]\right)\label{eq:W_sigma}
\end{align}
Thus, setting $r=1$ and $q=1$ yields the original SPICE formulation. Note that more general regularization functions could also be used, but in this paper, we confine our attention to the $\{r,q\}$-norm case, which we hereafter term the $\{r,q\}$-SPICE formulation. It should be noted that using an approach reminiscent of the one presented in \cite{Chartrand07_14}, it is possible to also consider the case when all $0<r,q<1$ resulting in a concave penalty term. Herein we restricted our attention to the case where $r\geq 1$ and $q\geq 1$. % However, this is outside the scope of this work.
%\begin{align}
%\underset{\{p_k\}}{\text{minimize}}\ \y^*\R^{-1}\y+\sum^{M}_{k=1}w_kp_k+\phi(\bsigma)
%\end{align}
%where $\phi(\cdot)$ is a regularizing function. In this work, however, we focus on the form \eqref{eq:newSPICE}. We coin the method for solving \eqref{eq:newSPICE} $q$-SPICE.
\section{Linking \{r,q\}-SPICE to penalized regression}\label{sec:SPICEtoLasso}
\noindent To demonstrate the effects of introducing the $r$- and the $q$-norm to SPICE, we follow the derivation in \cite{RojasKH13_61,StoicaZL14_33}, and proceed to examine the connection between $\{r,q\}$-SPICE and a penalized regression problem such as the LASSO expression in \eqref{eq:LASSO}. In order to do so, we distinguish between two cases, namely the case when each $\sigma_k$ is allowed to have a distinct value, and the case when all $\sigma_k$ are equal. First, we recall a lemma that will be helpful for the following derivation (see also \cite{StoicaZL14_33}).
\begin{algorithm} [t!]
\caption{The $\{r,q\}$-SPICE estimator with $r=1$} 
\label{alg:1} 
\begin{algorithmic} [1]
\STATE Initiate $p_k^{(0)} = |\bb_k^*\y|^2/||\bb_k||^4, \text{for } k=1,\dots,M$, $\sigma_k^{(0)}=|y_k|, \text{for } k=1,\dots,N$, and set $i=1$
\WHILE{the termination criteria is not fulfilled} 
\STATE Let $\R^{(i)} = \A\bP^{(i)}\A^*$
\STATE Form $\lambda$ from \eqref{eq:lambdaUpdate}
\STATE Update $p_k^{(i)}$ from \eqref{eq:pUpdate},  for each $k=1,\dots,M$
\STATE Update $\sigma_k^{(i)}$ from \eqref{eq:sigmaUpdate}, for each $k= 1,\dots,N$
\STATE Set $i = i+1$
\ENDWHILE
\end{algorithmic} 
\end{algorithm} 
\begin{theorem}
Let
\begin{align}
\tilde\bP = \textnormal{diag}\left(\left[\begin{array}{c c c} p_1 & \dots & p_M 
\end{array} \right]\right)
\end{align}
and
\begin{align}
\mathbf{\Sigma}= \textnormal{diag}\left(\left[\begin{array}{c c c} \sigma_1 & \dots & \sigma_N 
\end{array} \right]\right)
\end{align}
Then,
\begin{align}
\y^*\R^{-1}\y=\underset{\x}{\textnormal{minimize}}\ &(\y-\bB\x)^*\mathbf{\Sigma}^{-1}(\y-\bB\x)\nonumber\\&+\sum_{k=1}^{M}|x_k|^2/p_k
\end{align}
with the minimum occurring at
\begin{align}\label{eq:ptox}
\hat{\x}= \mathbf{\Sigma}\bB^*\R^{-1}\y
\end{align}\qed
\end{theorem}
\subsection{Varying noise variance}\label{subsectionA}
\noindent Using Lemma 1, one may rewrite \eqref{eq:newSPICE} as
\begin{align}\label{eq:reform}
\underset{\x,\bp, \sigma}{\text{minimize}}\ &\sum^N_{k=1}|y_k-\bb_k^*\x|^2/\sigma_k+\sum^M_{k=1}|x_k|^2/p_k\nonumber\\&+\left(\sum^{M}_{k=1}w_k^rp_k^r\right)^{1/r} +\left(\sum^{N}_{k=1}w_{M+k}^q\sigma_k^q \right)^{1/q}
\end{align}
\begin{figure}[t]
\vspace{-.5cm}
\includegraphics[width=3.7in,height = 2.5in]{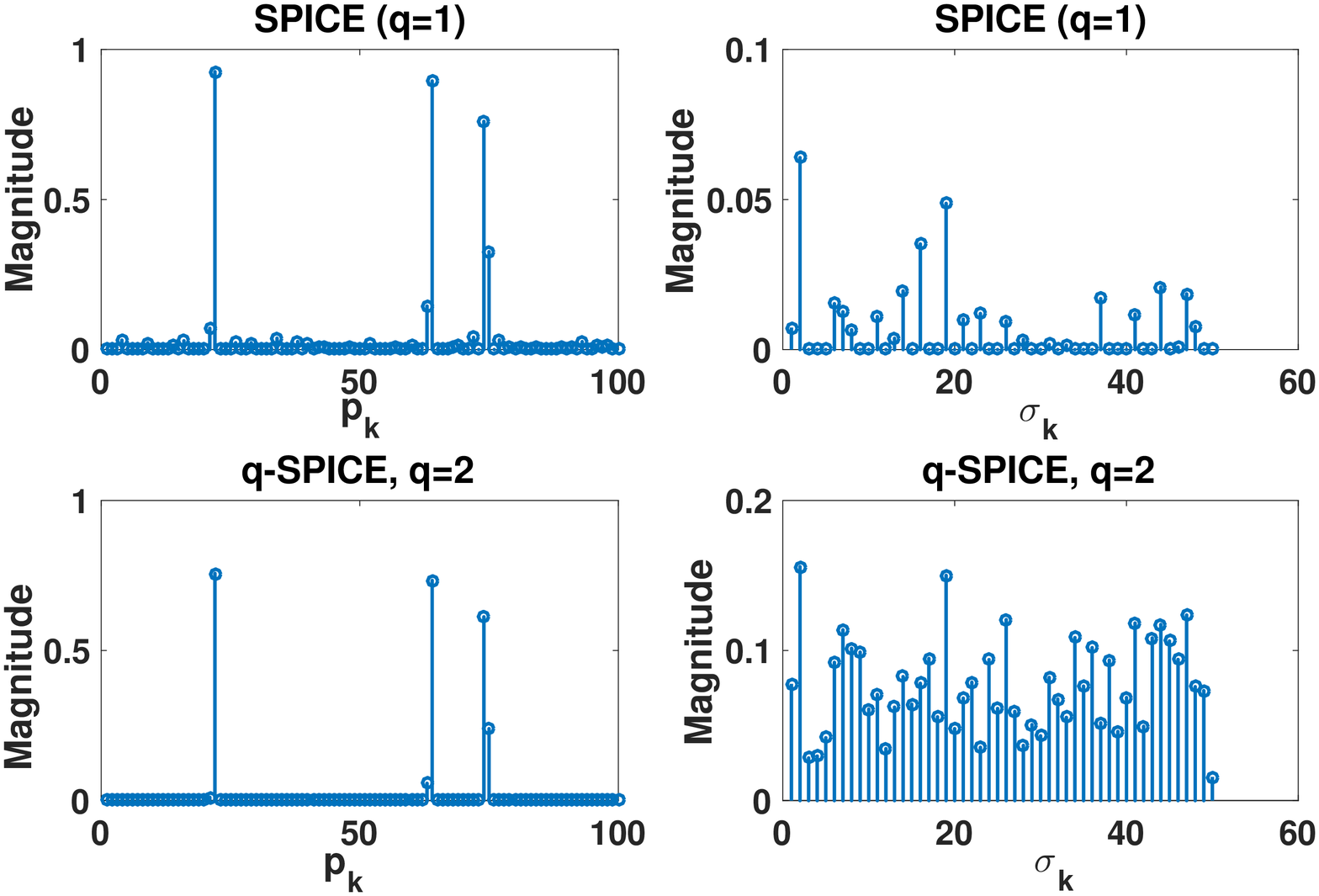} 
\caption{The resulting estimates of $\tilde\bp$ and $\mathbf{\sigma}$ from the SPICE and the $q$-SPICE estimator ($q$=2). Note that $q$-SPICE is sparser in $\tilde \bp$, whereas SPICE is sparser in $\mathbf{\sigma}$. In this example $r$ is set to $r=1$.} \label{fig:differenceSPICE_qSPICE}
\end{figure}
Solving \eqref{eq:reform} for $p_j$ yields
\begin{align}\label{eq:p}
p_j=w_k^{-\frac{r}{r+1}}|x_k|^{\frac{2}{r+1}}||\W^{1/2}\x||^{\frac{r-1}{r+1}}_{\frac{2r}{r+1}}
\end{align}
Differentiating the function to be minimized in \eqref{eq:reform} with respect to $\sigma_k$ and setting it to zero yields
\begin{align}\label{eq:diffSigma}
-\frac{|y_k-\bb_k^*\x|^2}{\sigma_k^2}+\frac{w_{M+k}^q\sigma_k^{q-1}}{||\W_\sigma\bsigma||_q^{q-1}}=0
\end{align}
Summing over $k$ on both sides and simplifying, one arrives at
\begin{align}\label{eq:Wsimple}
||\W_\sigma\bsigma||_q=||\W_\sigma^{1/2}\br||_{\frac{2q}{q+1}}
\end{align}
Inserting \eqref{eq:Wsimple} into \eqref{eq:diffSigma} yields
\begin{align}\label{eq:sigma}
\sigma_k = w_{M+k}^{-\frac{q}{q+1}}\left|r_k\right|^{\frac{2}{q+1}}\left|\left|\W_\sigma^{1/2}\br \right|\right|^{\frac{q-1}{q+1}}_{\frac{2q}{q+1}}
\end{align}
Finally, inserting \eqref{eq:p} and \eqref{eq:sigma} into \eqref{eq:reform} yields
\begin{align}\label{eq:penReg}
\underset{x}{\text{minimize}} \ \left|\left|\W_\sigma^{1/2}\left(\y-\bB\x\right)\right|\right|_{\frac{2q}{q+1}}+\left|\left|\W^{1/2}\x \right|\right|_{\frac{2r}{r+1}}
\end{align}
%where $\W$ is formed similar to \eqref{eq:W_sigma}, but using $w_k$ for $k=1,\dots,M$.
From the resulting expression, it may be noted that using $q=1$ yields the least absolute deviations (LAD) estimate, whereas using $q=\infty$ yields the (unscaled) square-root LASSO. The implications of this is discussed further below.
%\textcolor{red}{We should have a discussion on how $q$ affects the sparsity.}

\begin{algorithm} [t!]
\caption{The $\{r,q\}$-SPICE estimator for equal $\sigma_k$ with $r=1$.} 
\label{alg:2} 
\begin{algorithmic} [1]
\STATE Initiate $p_k^{(0)} = |\bb_k^*\y|^2/||\bb_k||^4, \text{for } k=1,\dots,M$, $\sigma^{(0)}=\sqrt{\frac{1}{N-1}\sum_{k=1}^N \left( y_k-\bar{y}\right)^2} \ , \text{for } k=1,\dots,N$, and set $i=1$
\WHILE{the termination criteria is not fulfilled} 
\STATE Let $\R^{(i)} = \A\bP^{(i)}\A^*$
\STATE Form $\lambda$ from \eqref{eq:lambdaUpdate2}
\STATE Update $p_k^{(i)}$ from \eqref{eq:pUpdate2},  for each $k=1,\dots,M$
\STATE Update $\sigma_k^{(i)}$ from \eqref{eq:sigmaUpdate2}, for each $k= 1,\dots,N$
\STATE Set $i = i+1$
\ENDWHILE
\end{algorithmic} 
\end{algorithm} 
Clearly, regardless of the choice of $q$, the corresponding problem in \eqref{eq:newSPICE} will still be scale invariant. To see this, we follow \cite{StoicaZL14_33} and scale each $p_k$ and $\sigma_k$ with a constant $c$ and do the same for the cost function in \eqref{eq:newSPICE}, defining 
\begin{align}
g(\bp,\bsigma) &\define c\y^*\left(\A c\bP \A^* \right)^{-1}\y \nonumber \\
&\ +c\left[ \sum^{M}_{k=1}w_k^rc^rp_k^r\right]^{1/r}+c\left[ \sum^{N+M}_{k=M+1}w_k^qc^qp_k^q\right]^{1/q}\nonumber \\
&=\y^*\left(\A\bP \A^* \right)^{-1}\y +c^2\left[ \sum^{M}_{k=1}w_k^rp_k^r\right]^{1/r}+\nonumber \\&\ c^2\left[ \sum^{N+M}_{k=M+1}w_k^qp_k^q\right]^{1/q}
\end{align}

\noindent
Defining the cost function in \eqref{eq:newSPICE} as $f(\bp,\bsigma)$, we may use  Lemma 2 in \cite{StoicaZL14_33} to conclude that if
\begin{align}
\{\hat\bp,\hat \bsigma \} = \underset{\bp,\bsigma}{\text{arg min}}\ g(\bp,\bsigma)
\end{align}
and
\begin{align}
\{\hat{\bar{\bp}},\hat{\bar{\bsigma}} \} =  \underset{\bar{\bp},\bar{\bsigma}}{\text{arg min}}\ f(\bar{\bp},\bar{\bsigma})
\end{align}
then 
\begin{align}
\hat{\bar{\bp}} = c\hat{\bp}
\end{align}
\noindent where $c>0$, which is true in the here examined case as well. Due to this scale invariance, one may conclude that the $\{r,q\}$-SPICE method is hyper-parameter free in the same sense as SPICE. Furthermore, it may be noted that when converting the $p_k$ to $x_k$, {using \eqref{eq:ptox}}, any scaling will disappear.

\subsection{Uniform noise  variance}\label{sec:sameSigma}
%If, similar to \cite{RojasKH13_61,StoicaZL14_33}, one instead assumes that all the noise terms have equal variance, thus treating $\sigma_k=\sigma,\ \forall k$, one arrives at an interesting conclusion: with this assumption, it has been shown \cite{RojasKH13_61,StoicaZL14_33} that the SPICE problem is equal to the square-root LASSO problem, i.e.,
%\begin{align}\label{eq:sqrtLASSO}
%\underset{\x}{\text{minimize}}\ ||\y-\bB\x||_2+\mu||\W^{1/2}\x||_1
%\end{align}
%where $\mu=N^{-1/2}$ for SPICE. Following the derivation in Secction \ref{subsectionA}, together with the assumption that all the noise terms have equal variance, one instead gets $\mu=N^{-1/2q}$.
%This means that increasing $q$ corresponds to increasing the sparsity in the square-root LASSO. This also indicates that for $q$-SPICE, in the case when $\sigma_k$ are allowed to differ, one should obtain sparser solutions with increasing $q$. Indeed, if choosing $q=\infty$, one obtains the square-root LASSO with $\mu=1$. This rational is confirmed in the numerical section below. 

\begin{figure}[t]
\vspace{-.5cm}
\includegraphics[width=3.7in,height = 2.5in]{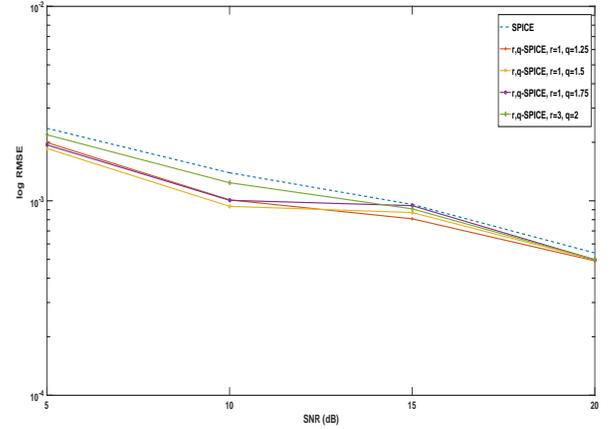} 
\caption{The RMSE of the frequency estimates, as a function of  SNR for $\{r,q\}$-SPICE and SPICE.} \label{fig:RMSE_gridless}
\end{figure}

%From the discussion above, it is clear that if one is able to find an efficient implementation of the $q$-SPICE approach, one may thus also efficiently solve the square-root LASSO.
\noindent
If, similar to \cite{RojasKH13_61,StoicaZL14_33}, one instead assumes that all the noise terms have equal variance, thus treating $\sigma_k=\sigma,\ \forall k$, one arrives at an interesting conclusion: with this assumption, it has been shown that the SPICE problem is connected to the (weighted) square-root LASSO problem \cite{RojasKH13_61,StoicaZL14_33}, i.e.,
\begin{align}\label{eq:sqrtLASSO}
\underset{\x}{\text{minimize}}\ ||\y-\bB\x||_2+\mu||\W^{1/2}\x||_1
\end{align}
where $\mu=N^{-1/2}$ yields the SPICE estimator. Following the derivation in Section \ref{subsectionA}, together with the assumption that all the noise terms have equal variance, yields $\mu=N^{-1/2q}$ for the $\{r,q\}$-SPICE formulation, implying the equivalent formulation 
\begin{align}\label{eq:sameSigmaPenReg}
\underset{\x}{\text{minimize}}\ ||\y-\bB\x||_2+\mu||\W^{1/2}\x||_{\frac{2r}{r+1}}
\end{align}
Thus, the choice of $q$ corresponds to the weight that governs the trade-off between the model fitting term and the regularization of the parameters, and the choice of $r$ decides which norm will be used in the regularization of the parameters. Thus, using $r=1$ means that increasing $q$ corresponds to increasing the sparsity in the (weighted) square-root LASSO{; this implies that if the signal at hand is assumed to be sparse, solving $\{r,q,\}$-SPICE with $q>1$ will yield preferable estimates.} Furthermore, setting $r\rightarrow\infty$ yields a ridge regression problem, with $q$ governing the amount of regularization. We note that it might be preferable to solve \eqref{eq:sameSigmaPenReg} using the $\{r,q\}$-SPICE formulation, rather than solving \eqref{eq:sameSigmaPenReg} directly.

\section{Efficient implementation}
\noindent
As will be argued later, for sparse problems, the most interesting setting for $\{r,q\}$-SPICE is when $r=1$, since, according to \eqref{eq:sameSigmaPenReg}, this will yield an $\ell_1$ regularization. To this end, we will in this section derive an efficient implementation for this case. In \cite{StoicaBL11_59}, an efficient implementation of SPICE was introduced. To derive the steps of this algorithm, it was noted that the original SPICE minimization in \eqref{eq:SPICE} could also be expressed as
\begin{figure}[t]
\includegraphics[width=3.7in,height = 2.5in]{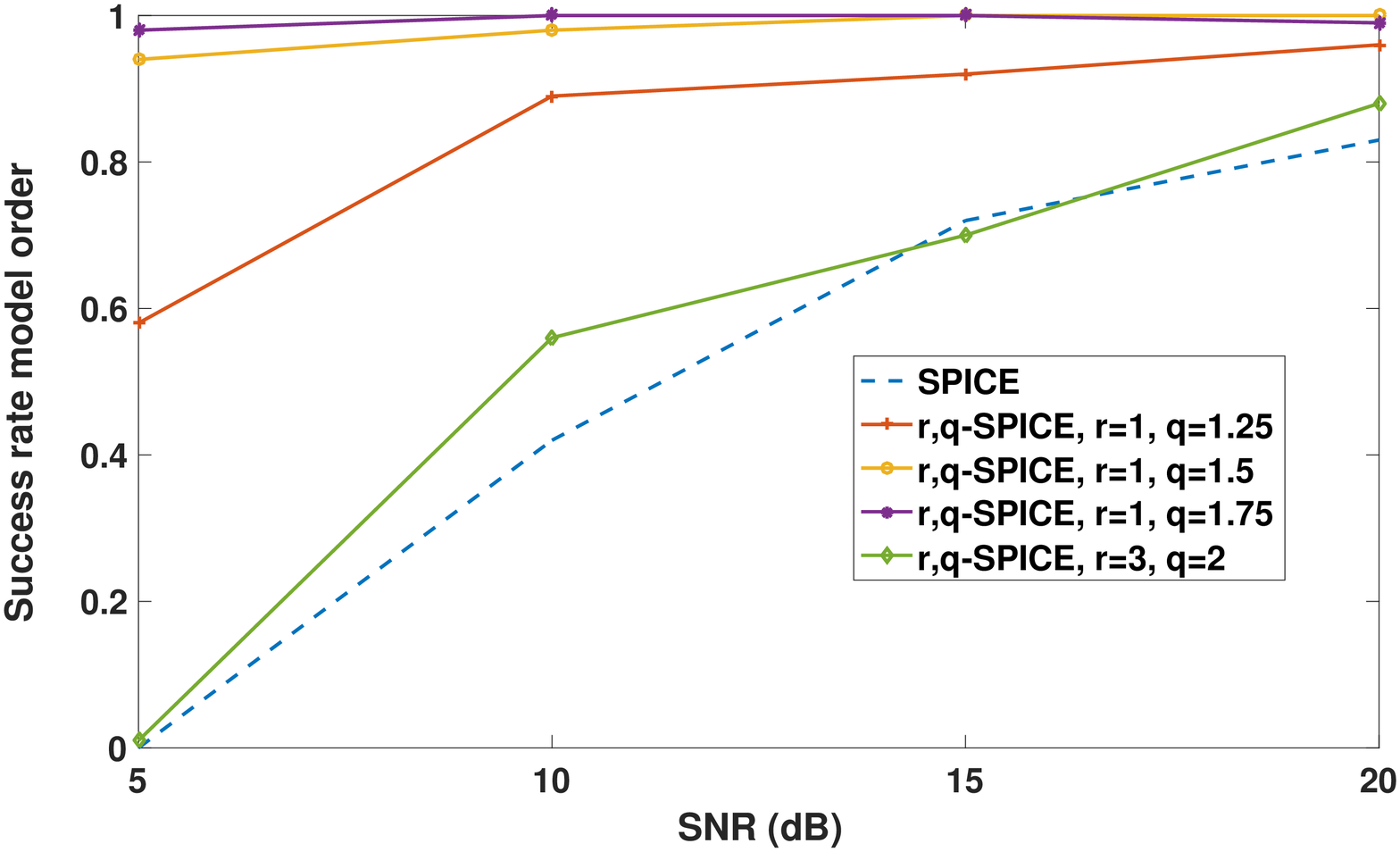} 
\caption{The probability of finding the correct model order of the signal as a function of SNR for $\{r,q\}$-SPICE and SPICE.} \label{fig:modelOrder_gridlessRQSPICE}
\end{figure}
\noindent
\begin{align}\label{eq:SPICE2}
\underset{\{p_k\geq 0\}_{k=1}^M, \ {\{\sigma_k\geq0\}_{k=1}^N}}{\text{minimize}}\ \y^*\R^{-1}\y \ \text{subject to}\ 
\end{align}
Furthermore, it was noted that one could further rewrite the objective in \eqref{eq:SPICE2} by considering the optimization problem
\begin{align}
\underset{\bQ}{\text{minimize}}\ \y^*\bQ^*\bP^{-1}\bQ\y\ \text{subject to}\ \bQ^* \A=\mathbf{I}
\end{align}
which has the solution $\bQ_0 = \bP\A^*\R^{-1}$. By defining 
\begin{align}\label{eq:beta}
\boldsymbol{\beta} = \bQ\y
\end{align}
one may rewrite \eqref{eq:SPICE2} as
\begin{align} \label{eq:betaSPICE}
\underset{\{p_k\geq 0\}_{k=1}^M,\ {\{\sigma_k\geq0\}_{k=1}^N}}{\text{minimize}}\ \sum^{M+N}_{k=1}\frac{|\beta_k|^2}{p_k}\ \ \text{subject to}\ \ {\sum^{M}_{k=1}w_kp_k+\sum^N_{k=1}w_k\sigma_k=1}
\end{align}
The estimates may then be found by iteratively updating $\R$ and solving for $p_k$ in \eqref{eq:betaSPICE}. For $\{r,q\}$-SPICE, with $r=1$, when assuming different values for the $\sigma_k$, the same update for $\R$ may be used, but instead of \eqref{eq:betaSPICE}, one needs to solve

\begin{figure}[t]
\includegraphics[width=3.7in,height = 2.5in]{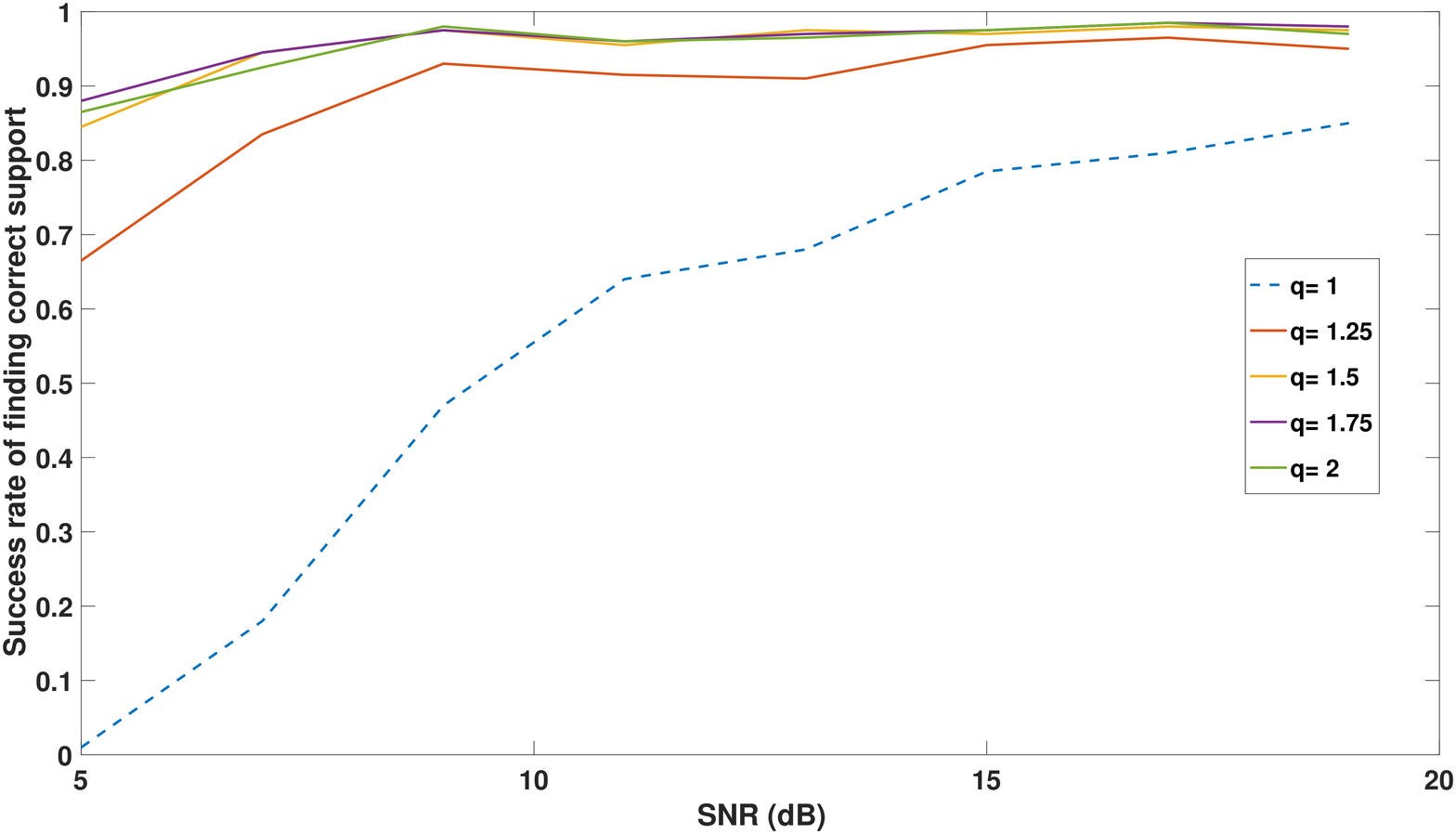} 
\caption{The probability of finding the correct support of the signal as a function of $q$ and SNR. Here, all the $\sigma_k$ are assumed to be equal. In this example, $r=1$.} \label{fig:q1Toq3}
\end{figure}

\begin{align}
&\underset{\{p_k\geq0\}_{k=1}^M,\{\sigma_k\geq0\}_{k=1}^N}{\text{minimize}}\ \sum^{M}_{k=1}\frac{|\beta_k|^2}{p_k}+\sum^N_{k=1}\frac{|\beta_{M+k}|^2}{\sigma_k}\nonumber\\  &\text{subject to}\ \sum^{M+N}_{k=1}w_k p_k+\left(\sum^N_{k=1}w_{M+k}^q\sigma_k^q \right)^{1/q}=1
\end{align}
From the Karush-Kuhn-Tucker (KKT) conditions \cite{BoydV04}, it follows that
\begin{align}
&-\frac{|\beta_k|^2}{p_k^2}+\lambda w_k=0,\ \text{for } k=1,\dots,M \\
&-\frac{|\beta_{M+k}|^2}{\sigma_k^2}+\lambda\sigma_k^q w^q_{M+k}\left(\sum^N_{k=1}w_{M+k}^q\sigma_k^{q-1}\right)^{1/q}=0
\end{align}
where $\lambda$ denotes the dual variable, for $k=1,\dots,M$, together with the constraint in \eqref{eq:betaSPICE}. Solving these equation for each $p_k$ and $\sigma_k$ yields
\begin{align}\label{eq:pUpdate}
p_k &= \frac{|\beta_k|}{\sqrt{w_k}\lambda^{1/2}}\\ 
\label{eq:sigmaUpdate}
\sigma_\ell &=\frac{|\beta_{M+\ell}|^{\frac{2}{q+1}}||\W_\sigma^{1/2}\boldsymbol{\beta}_\sigma||_{\frac{2q}{q+1}}^{\frac{q-1}{q+1}}}{w_{M+\ell}^\frac{q}{q+1}\lambda^{1/2}} \\
 \label{eq:lambdaUpdate}
\lambda&= \left(||\W^{1/2}\boldsymbol{\beta}||_1+||\W_\sigma^{1/2}\boldsymbol{\beta}_\sigma||_{\frac{2q}{q+1}} \right)^2
\end{align}
for $k=1,\dots,M$ and $\ell = 1,\dots,N$, where 
\begin{align}
\boldsymbol{\beta}&=\left[\begin{array}{c c c} \beta_1 & \dots & \beta_M\end{array}\right]^T\\
\boldsymbol{\beta}_\sigma&=\left[\begin{array}{c c c} \beta_{M+1} & \dots & \beta_{M+N}\end{array}\right]^T
\end{align}
This allows for the formulation of an efficient implementation by iteratively forming $\R$ from \eqref{eq:R}, $\beta_k$ from \eqref{eq:beta}, and $p_k$ and $\sigma_k$ from \eqref{eq:pUpdate} and \eqref{eq:sigmaUpdate}, respectively. Since $\{1,q\}$-SPICE allows for a more sparse solution than the original SPICE, one may speed up the computations further by removing the zero valued $p_k$ when forming $\R$ and $\beta_k$.

\begin{figure}[t]
\includegraphics[width=3.7in,height = 2.5in]{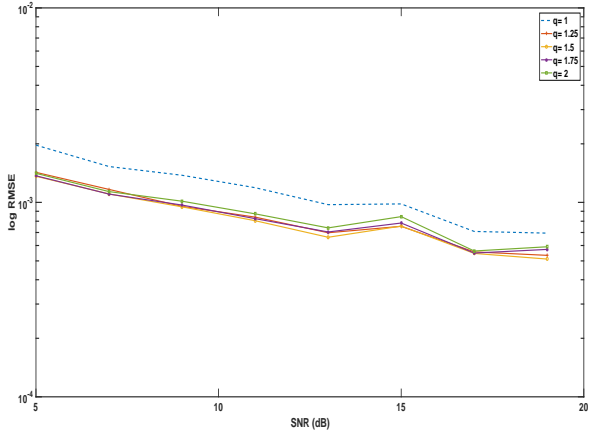} 
\caption{The RMSE  of the frequency estimates, as a function of $q$ and SNR. Here, all the $\sigma_k$ are assumed to be equal. In this example, $r=1$.} \label{fig:RMSE_q1_q5}
\end{figure}

When instead assuming that $\sigma_k=\sigma,\ \forall k$, one obtains the steps
\begin{align}\label{eq:pUpdate2}
p_k &= \frac{|\beta_k|}{\sqrt{w_k}\lambda^{1/2}}\\ \label{eq:sigmaUpdate2}
\sigma &=\frac{||\beta_M||_2}{N^{1/2q}\lambda^{1/2}} \\ \label{eq:lambdaUpdate2}
\lambda&= \left( ||\W^{1/2}\boldsymbol{\beta}||_1+||N^{1/(2q)}\boldsymbol{\beta}_\sigma||_2 \right)^2
\end{align}
for $k = 1,\dots,M$. Algorithms \ref{alg:1} and \ref{alg:2} summarize the $\{1,q\}$-SPICE implementations for the two settings, with $\bar{y}$ denoting the mean value of the vector $\y$. Similar to the previous case, since using $q>1$ will enforce more sparsity than $q=1$, one may utilize this added sparsity in the implementation of the algorithm. 
Since most of the elements in $\bp$ will be zero, one may form $\R^{-1}$ by only considering the columns and rows of $\A$ and $\A^*$ corresponding to the non-zero entries in $\bp$. Let $\hat{K}^{(i)}$ be the number of non-zero entries in $\bp^{(i)}$ at iteration $i$. Then, if $\hat{K}< N$, one may use the Woodbury matrix identity to efficiently calculate the inverse of $\R$ (see, e.g., \cite{GolubV13}).

The termination criterias in Algorithms \ref{alg:1} and \ref{alg:2} can take on many forms. In this work, we have chosen to terminate the algorithms when the percentage of change in $\bp$ and $\boldsymbol{\sigma}$ between two consecutive iterations falls below a certain level, say in the range $[10^{-9}, 10^{-3}]$. %Other stopping criteria include a fix number of iterations, rate of change in the cost function, or some information theoretical criteria.

Note that the algorithm described in Algorithm \ref{alg:2} solves a (weighted) square-root LASSO problem, where the different choices of $q$ corresponds to different levels of sparsity, i.e., different values of $\mu$ in \eqref{eq:sqrtLASSO}.  If one is interested in solving a (weighted) square-root LASSO with $\mu=\mu_0$, then one may instead solve the $\{r,q\}$-SPICE with $q=-\frac{1}{2\ln \mu_0}$, as long as $q>1$, and with $r=1$.
Thus, the algorithm in Algorithm~\ref{alg:2} presents an attractive and efficient way of solving the (weighted) square-root LASSO problem, for a large range of different $\mu$. 

To give an idea of the running time of the proposed algorithm as compared with a standard SDP solver (see, e.g., \cite{cvx,GrantB08}), the algorithms were tested on a problem with $M=10000$,  $N=1000$, and with $q=5$, and $r=1$, where the data vector, $\y$, contained $3$ sinusoids, using a standard PC (2.6 Ghz Intel Core i7, 16 GB RAM). The corresponding run times were roughly $4$ seconds for the Matlab implementation in Algorithm~\ref{alg:2} and $ 4132$ seconds for the SDP Matlab solver\footnote{Our implementation of $\{r,q\}$-SPICE will be made available on the authors' web-pages upon publication.}.

\section{Off-grid solution}
%Solving estimation problems by evaluating the solution by forming a grid over the parameters is an approach that has been practiced for years. Lately, however, a couple of concerns with this approach have been raised. For instance, what happens when the true parameter values lies off the grid? 
\noindent
Many forms of estimation problems are solved by evaluating over a grid of the parameters of interest. However, such a solution may cause concerns when the sought solution falls outside the grid or may be found in between grid points. A common solution to this problem is to increase the grid size to thereby minimize the distance from the closest grid point to the true parameter value (see, e.g., \cite{StoicaB12_60,ChiSPC11_59}). However, such a solution might cause the columns of the extended dictionary to be highly correlated, thereby decreasing the performance of the method (we instead refer the interested reader to other works treating this issue, e.g., \cite{ChiSPC11_59,TangBSR13_59,BhaskarTR13_61,YangXZ12_61} and the references therein). In \cite{YangX15_63} and \cite{StoicaTYZ14_eusipco}, an off-grid solution to the original SPICE version was presented for the sinusoidal case.
\begin{figure}[t]
\includegraphics[width=3.7in,height = 2.5in]{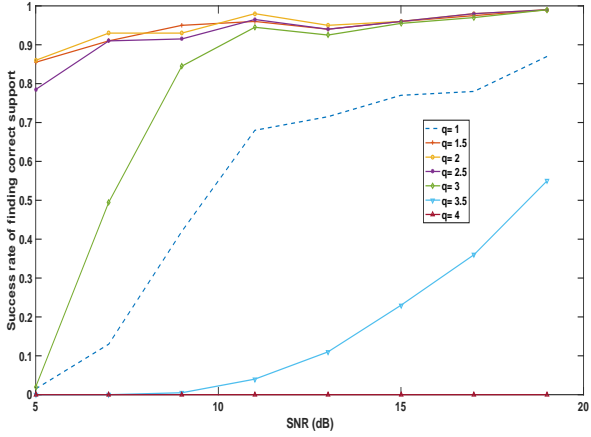} 
\caption{The probability of finding the correct support of the signal as a function of $q$ and SNR. Here, all the $\sigma_k$ are assumed to be equal. Here, $r=1$.} \label{fig:q1Toq9}
\end{figure}
In this section, we similarly provide one possible version of off-grid estimation for the proposed $\{r,q\}$-SPICE method for a signal containing superimposed sinusoids. In order to do so, it may initially be noted that one  may separate $\R$ into two different matrices, such that
\begin{align}
\R = \bB^*\text{diag}\left( {\bp}\right)\bB + \text{diag}\left( \bsigma\right)\define \bT(\bu) +\text{diag}\left( \bsigma\right)
\end{align}
where $\bT(\bu)$ is a Toeplitz matrix with $\bu$ forming the first column of $\bT(\bu)$.
Thus, \eqref{eq:newSPICE} may be expressed as {(see also  \cite{YangX15_63,StoicaTYZ14_eusipco})}
\begin{align}\label{eq:offGridSPICE}\nonumber
\underset{ \bu,\bsigma,x}{\text{minimize}}\ & ||\y||_2^2x+||\text{diag}(\bT(\bu))||_r+||\W_\sigma \bsigma||_q&\\
\text{subject to} \ &\begin{bmatrix}
x & \y^* \\ \y & \bT(\bu) +\text{diag}\left( \bsigma\right) \nonumber
\end{bmatrix}\geq 0\\ 
&\nonumber \bT(\bu) \geq 0 \\ \nonumber
&\bT(\bu) - \bT(\bu)^* = 0\\
&\bsigma \geq 0
\end{align}
and under the additional constraint that $\bT(\bu)$ is a Toeplitz matrix. The optimization problem in \eqref{eq:offGridSPICE} is convex, and may be solved using, e.g., a publicly available SDP solver, such as the one presented in \cite{cvx,GrantB08}. The final off-grid estimates may then be found using the celebrated Vandermonde decomposition in combination with, for instance, Prony's method (see \cite{BluDVMC08_25,StoicaM05} for further details on such an approach).

\begin{figure}[t]
\includegraphics[width=3.7in,height = 2.5in]{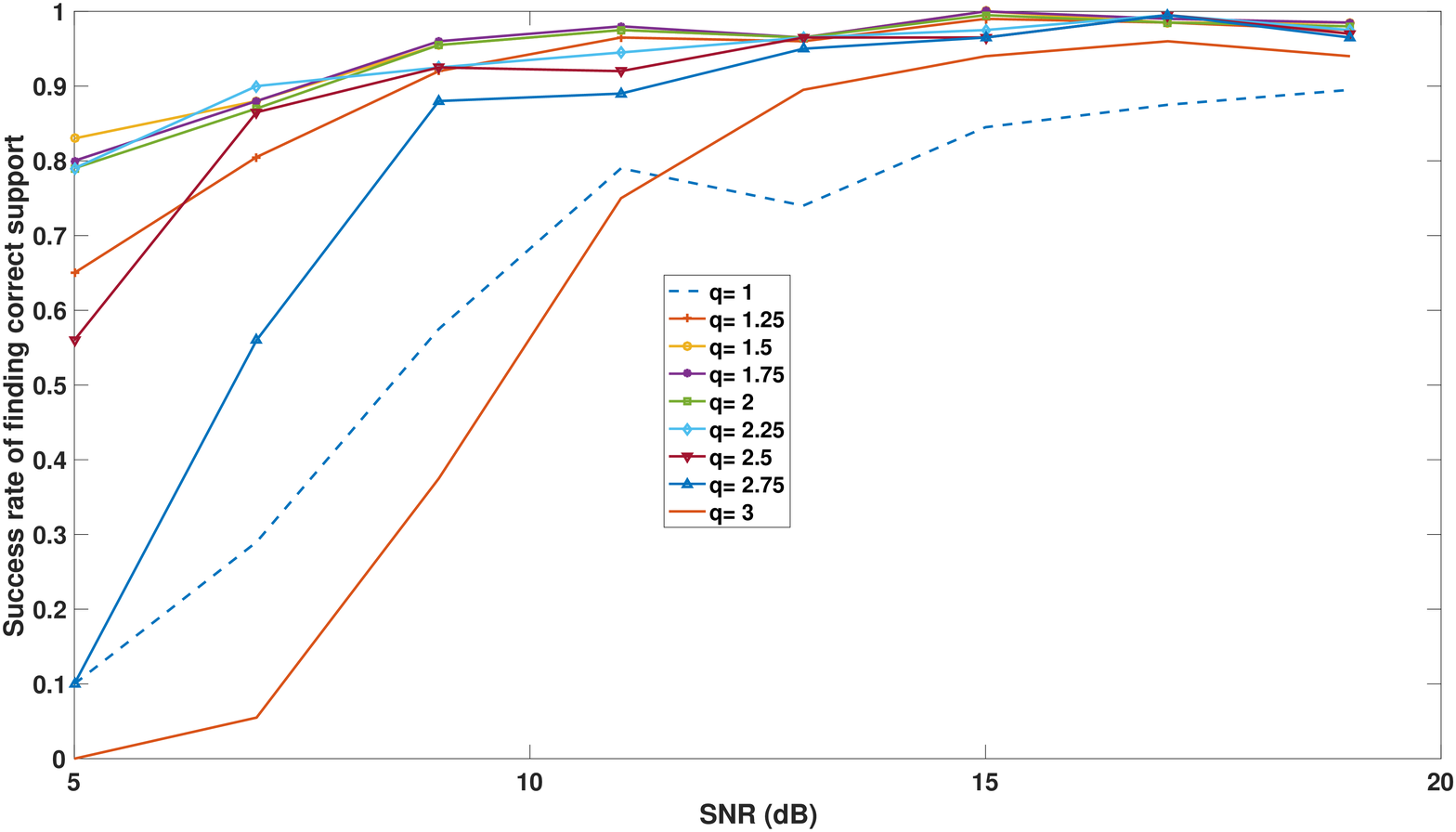} 
\caption{The probability of finding the correct support of the signal as a function of $q$ and SNR. Here, $r=1$.} \label{fig:q1_q3differentSigmas}
\end{figure}

\section{Numerical examples}
\noindent
Using the interpretation provided by the reformulation in Section \ref{sec:SPICEtoLasso}, it is clear that the choice of $r$ will decide what kind of regularization that will be used. Thus, choosing $r=1$ will yield an $\ell_1$ norm and letting $r\rightarrow \infty$ will result in the $\ell_2$ norm. In this paper, we consider sparse problems, and will therefore mainly confine our attention to the case where $r=1$, since this will yield the most sparse convex regularizer, namely $\ell_1$. 

From the discussion in Section \ref{sec:II}, one may expect that SPICE will set some of the elements in $\bsigma$ to zero, since the sparsity enforcing term in \eqref{eq:SPICE} also applies to these parameters. Figure~\ref{fig:differenceSPICE_qSPICE} shows the estimated $\bp$ and $\bsigma$ for the SPICE and the $\{r,q\}$-SPICE estimators, when applied to a linear signal formed using \eqref{eq:model} with three non-zero components. As expected, using $r=1$, $\{r,q\}$-SPICE offers a sparser $\bp$ vector as compared to SPICE, whereas the solution is more sparse in $\bsigma$ for SPICE. As a result, the sparsity constraints on the $\sigma_k$ terms in $\{r,q\}$-SPICE are thus relaxed and are instead subjected to a bounding of their power in the $q$-norm, thus allowing for more sparsity in $\bp$.

\begin{figure}[t]
\includegraphics[width=3.7in,height = 2.5in]{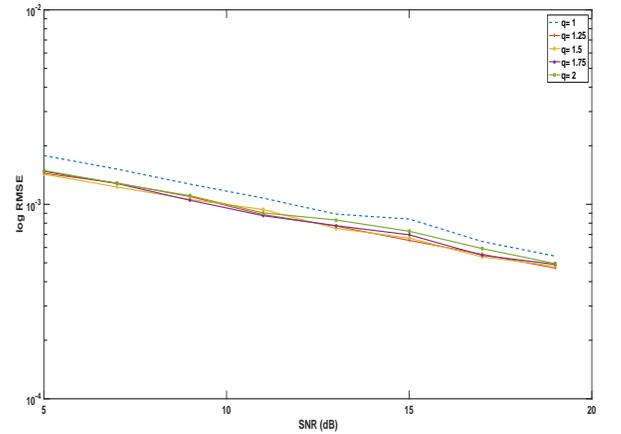} 
\caption{The RMSE of the frequency estimates, as a function of $q$ and SNR. Here, $r=1$.} \label{fig:RMSE_q1_q2d5differentSigmas}
\end{figure}

We will proceed by showing the difference in performance for different values of $r$ and $q$, to provide an example on how the different choices of these norms affect the estimates. We investigate two properties of the estimators, namely the resulting root-mean-squared error (RMSE) of the frequency estimates, defined as
\begin{align}\label{eq:RMSE}
\text{RMSE}\define\sqrt{\frac{1}{P}\sum_{k=1}^{P} |\hat \theta_k-\theta_k |^2}
\end{align}
\noindent where $\theta_k$ is the true frequency of the $k$th component, whereas $\hat\theta_k$ is the formed estimate, and the ability to correctly estimate the model order. The signal was $N=50$ samples long and contained $4$ sinusoids with unit magnitude and random phase. The simulation was done using $100$ Monte-Carlo simulations for each SNR-level, where the signal-to-noise ratio (SNR) is defined as 
\begin{align}
\text{SNR}=10\log_{10}\left(\frac{P_y}{\sigma^2}\right)
\end{align}
with $P_y$ denoting the power of the true signal. The noise used was circular white Gaussian noise, and the noise terms were allowed to differ. 

\begin{figure}[t]
\includegraphics[width=3.7in,height = 2.5in]{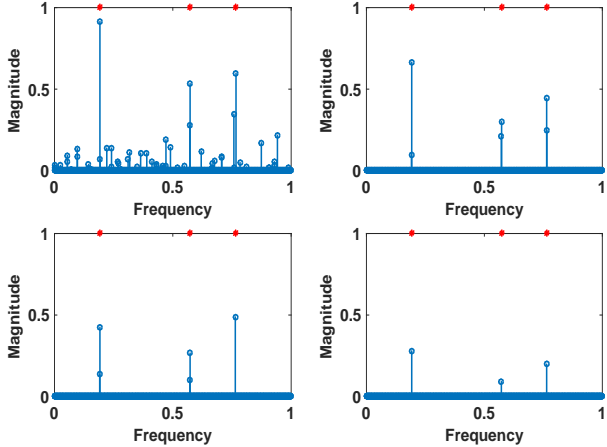} 
\caption{A typical result from $q$-SPICE for different values of $q$. Top left: $q=1$, top right $q=1.5$, bottom left $q=2$, and bottom right $q=2.5$. The red stars indicate the position and size of the true sinusoids. In this example, $r=1$.} \label{fig:comparison}
\end{figure}
The solution was obtained by solving \eqref{eq:offGridSPICE} for all settings except for the original SPICE, where the estimates were obtained from solving the problem formulated in \cite{StoicaTYZ14_eusipco}. In Figure \ref{fig:RMSE_gridless}, the resulting RMSEs are shown for different values of $r$ and $q$, as a function of the SNR. To make the figures readable, one respectively two outliers were removed for SPICE and for the $r=3, q=2$ case for $\{r,q\}$-SPICE at the $5$ dB SNR-level. Furthermore, to remove the noise peaks that appear when using small values of $q$, all peaks smaller than $20$ \% of the largest found peak were removed. Note, however, that this is not necessary for the case where $q$ is larger. 
As is clear from the figure, the RMSE is decreased as the sparsity level is increased, with the $\{r,q\}$-SPICE versions outperforming the original SPICE. This is also true for the resulting model order estimation, which is shown in Figure \ref{fig:modelOrder_gridlessRQSPICE}. As may be expected, when increasing $q$ the sparsity is increased and the spurious peaks are removed, but as $q$ is further increased, the true peaks start to disappear. In this setting, it seems to be beneficial to set the norms around $q=1.5$ and $r=1$. From these results, we conclude that the generalized version of SPICE allows for better estimation of parameter values, as well as model order. As was expected, using $r>1$ was not beneficial when confronted with a sparse signal, and we will therefore, in the succeeding example, restrict our attention to the case where $r=1$, referring to the method as $q$-SPICE. However, it should be stressed that for certain situations, it might be preferable to use $r>1$, e.g., in situations when otherwise considering to use ridge regression; we will further examine this aspect in future works.

Arguably, the most important property of a sparse estimator is the ability to return the true support of the signal, as well as yielding reasonable amplitude estimates for this support. However, it seems inevitable that when including a sparsity enforcing penalty, one also introduced a (downwards) bias on the magnitude of the amplitudes. Fortunately, this problem is often easy to overcome by simply re-estimating the amplitudes using, e.g., least squares, once the true support is known.
Accordingly, we will in this section focus on the methods ability of finding the true support of the signal. To this end, {$200$} Monte-Carlo simulation for each SNR level are formed.  In each simulation, $N=50$ samples of a signal containing three sinusoids, each with unit magnitude, and phase uniformly drawn from ${(}0,2\pi]$, was created. The normalized frequencies were uniformly selected, but were at least $1/2N$ apart. 
The dictionary contained $M=1000$ candidate sinusoids, selected on a uniform frequency grid from $(0,1]$. The estimated support was selected to be the elements of the vector $\x$ that had a corresponding absolute value of at least $20\%$ of the largest estimated value in $\x$. This was done to allow for comparison with the less sparse $q$-SPICE versions, for cases with small $q$ value (most notably $q=1$). It may be noted that for values of $q$ that are large, this is not necessary.  The support was deemed correctly estimated if the estimated frequencies were at most two grid points away from the true frequencies.
Figure~\ref{fig:q1Toq3} shows the results of applying $q$-SPICE, for different values of $q$, assuming that all the $\sigma_k$ are the same, with $q=1$ yielding the SPICE estimate. As is clear from the figure, the results improve with increasing $q$ values. From the discussion in Section~\ref{sec:sameSigma}, we note that this corresponds to increasing the value of $\mu$ in \eqref{eq:sqrtLASSO}, thus increasing the sparsity in the estimates. Thus, one could assume that when further increasing $q$, the estimate of the support should decline. In Figure~\ref{fig:q1Toq9}, this behavior can be seen, where now $q$-SPICE is evaluated over a range of larger $q$ values. 
It is also apparent from the figure that the best value for $q$ is for this signal somewhere around $q= 2$, which corresponds to using $\mu \approx{0.38}$ in {\eqref{eq:sameSigmaPenReg}}.
Next, we investigate the precision for different values of $q$, by using the RMSE of the frequency estimates. Figure~\ref{fig:RMSE_q1_q5} shows the resulting RMSE of the frequency estimates, for the three largest values of $\x$. As can be seen in the figure, the RMSE is clearly improving as $q$ is increased, corresponding to sparser solutions. For smaller values of $q$, the results are not very sparse, and large spurious noise peaks can be found. 
To improve readability, {seven, two, and three outliers were removed from the cases $q=1$, $q=1.25$, and $q=1.5$, respectively}. If $q$ is increased too much this will, of course, make the solution too sparse, thus risking setting non-noise peaks to zero. This can also be seen in Figure~\ref{fig:q1Toq9}, where for about $q= 3$, the probability of retrieving the true support of the signal starts to decline{, and at $q>3.5$, the solution is too sparse.}

\begin{figure}[t]
\includegraphics[width=3.7in,height = 2.5in]{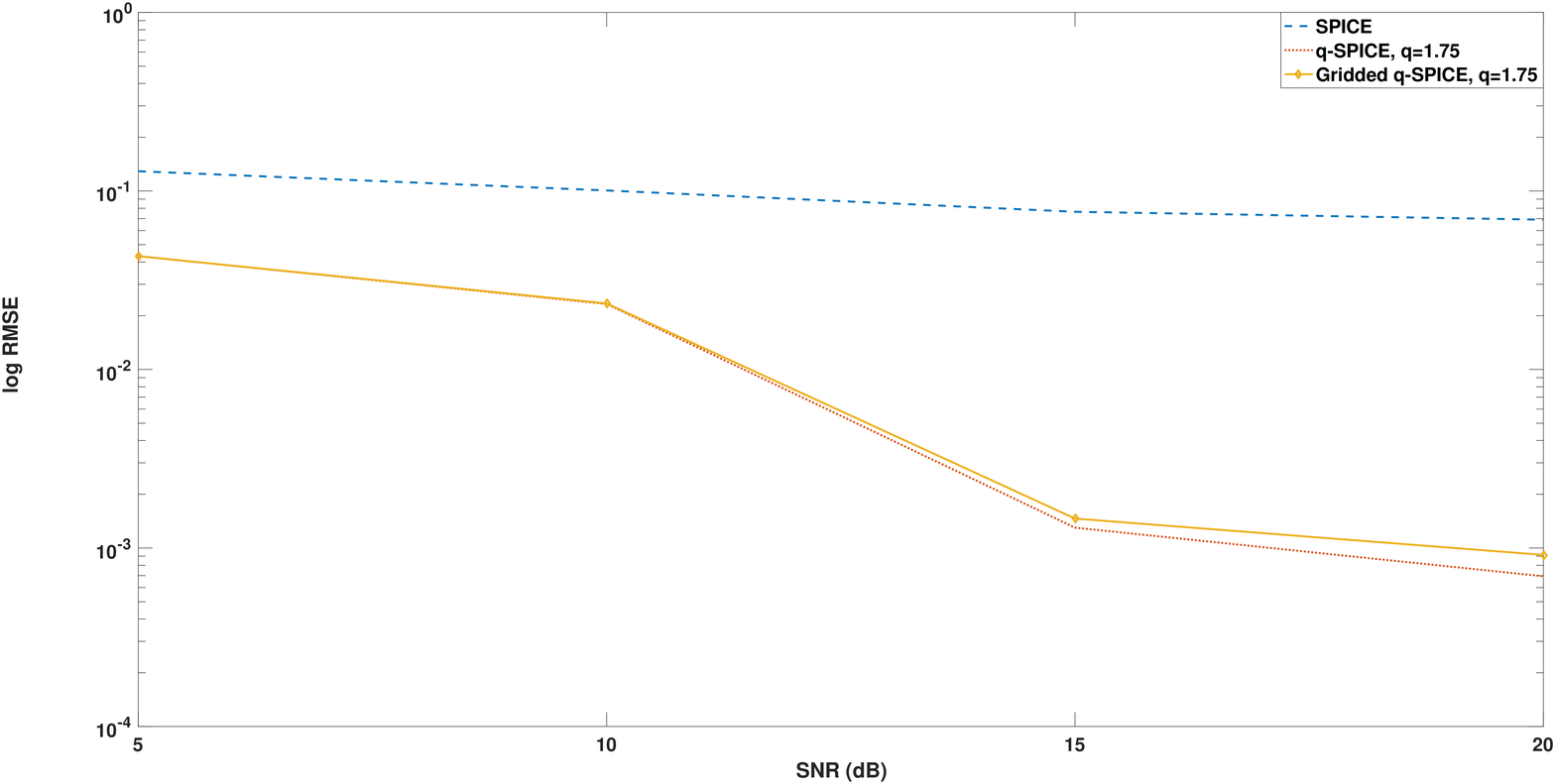} 
\caption{The RMSE of the frequency estimates, as defined in \eqref{eq:RMSE}, as a function of  SNR for the gridless versions of $q$-SPICE and SPICE, together with the gridded version of $q$-SPICE.} \label{fig:RMSE_Qgridless}
\end{figure}
We proceed by considering the case when the $\sigma_k$ parameters are allowed to take on different values, using the same set-up as above.
Figures \ref{fig:q1_q3differentSigmas} and \ref{fig:RMSE_q1_q2d5differentSigmas} show the probability of estimating the correct support of the signal and the RMSE of the three largest frequency estimates, respectively. Again, in the interest of readability, three outliers were removed from $q=1$, { six outliers from $q=1.25$, and three outliers for $q=1.5$}. As previously noted, it is clear from the figures that $q$ governs the sparsity enforced on the solution. From the figures, one may also see that for this setup, it is advantageous to choose $q$ in the interval {$q=[1.25,2.25]$}.
To demonstrate the differences in the solutions obtained from using different values of $q$, we show a typical simulation result for four different values of $q$, namely $q=1,1.5,2,$ and $2.5$, for the settings above, with SNR$=5$ dB. Figure~\ref{fig:comparison} shows the results, where it may again be noted that the sparsity level increases with $q$.

\begin{figure}[t]
\includegraphics[width=3.7in,height = 2.5in]{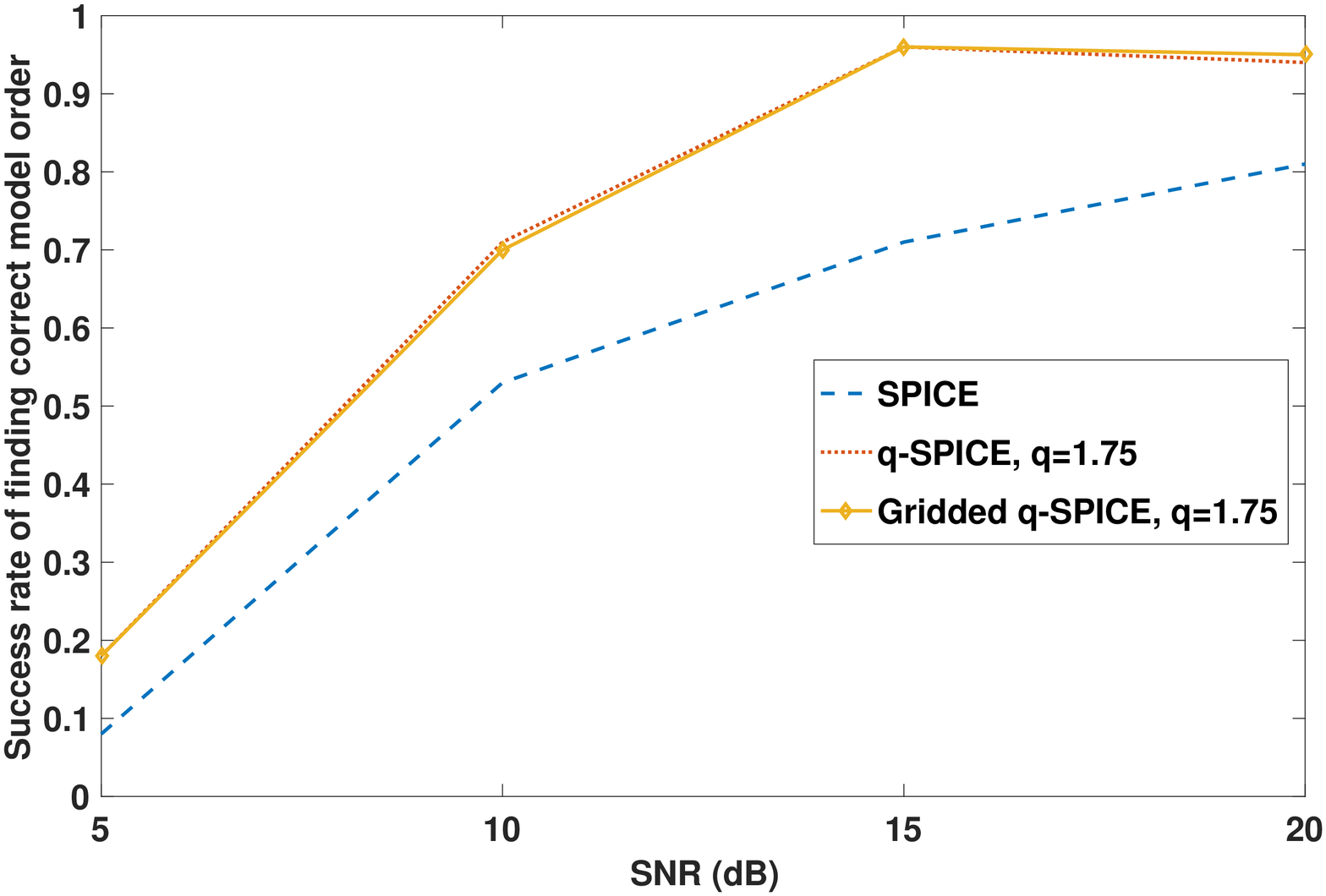} 
\caption{The probability of finding the correct model order of the signal as a function of  SNR for the gridless versions of $q$-SPICE and SPICE, together with the gridded version of $q$-SPICE.} \label{fig:modelOrder_gridlessQSPICE}
\end{figure}

Finally, we provide a numerical example showing the results from solving the $\{r,q\}$-SPICE using \eqref{eq:offGridSPICE}, with $r=1$ and $q=1.75$, and for the case where each noise variance are allowed to differ across the samples. In this scenario, we evaluated the gridless version of $\{r,q\}$-SPICE, given in \eqref{eq:offGridSPICE}, and the gridless version of SPICE, given in \cite{StoicaTYZ14_eusipco}, together with the grid-based $\{r,q\}$-SPICE, given a frequency grid of $M=500$ grid points. In each of the $100$ Monte-Carlo simulations, the $N=50$ samples long signal contained four sinusoids, each with random phase, with two peaks having magnitude $4$, one peak magnitude $2$, and the last one unit magnitude.
The frequencies were selected not to be closer than $1/2N$ from each other and were randomly selected in each simulation from the interval $(0,1]$. Both gridless versions were computed using the SDP-solver in CVX \cite{cvx,GrantB08}. Figure~\ref{fig:RMSE_Qgridless} and \ref{fig:modelOrder_gridlessQSPICE} show the resulting RMSE and probability of finding the correct support as functions of the SNR level. As seen in the figures, the two versions of the $q$-SPICE outperforms the gridless version of SPICE.
It is worth noting that in this scenario, only SPICE had the benefit of removing the smallest peaks. Furthermore, the model order was deemed correct if the method found the true number of peaks, thus there were no limitation on how close an estimated frequency had to be the true value. If the model order was too high, the four largest peaks were selected to compute the RMSE, whereas if the model order was too low, these estimates were omitted from the RMSE evaluations. 
Furthermore, one may see that the gridless version of $q$-SPICE is slightly better than the gridded version. However, this slight improvement from using the gridless $q$-SPICE version may not be worth the extra computation time; the gridless version took on average $9.4$ seconds to execute, whereas the gridded version only took $0.5$ seconds. However, it is worth recalling that other works on gridless solutions implicate that faster implementations are available (see, e.g.,\cite{YangX16_64}), and these improvements in implementation can likely also be applied to the gridless $q$-SPICE.

\section{Conclusion}
In this paper, we introduced a generalization of the SPICE method, in which we allow for a trade-off between the penalties for the model, using a $q$-norm, and the noise parameters, using an $r$-norm. We show that for larger values of $q$, one achieves a higher level of sparsity and better performance for recovering the support of the signal. Furthermore, we show that the proposed method is equivalent to a penalized regression formulation, with the $\frac{2q}{q+1}$ norm on the model fit, for the case when we let the noise variance vary across all samples. In the case where the noise variance is assumed to be equal for all samples, it is shown that the proposed method is equal to the (weighted) square-root LASSO, where the regularization parameter has a one-to-one correspondence to the choice of $q$ for a given problem. Furthermore, we provide a fast and efficient implementation for both the case when $r=1$ and the noise variances are equal for all samples, and where they are allowed to differ. As a result of the shown equivalence, the presented implementation offers an attractive alternative for solving $\frac{2q}{q+1}$-norm problems, and, perhaps more interesting, (weighted) square-root LASSO problems for different regularization parameters. We also present a gridless version of $\{r,q\}$-SPICE for the sinusoidal signals, which is on the form of an SDP problem. Numerical result show the preferred performance of the $\{r,q\}$-SPICE as compared to the original SPICE method, both for gridded and for gridless versions for the estimator.
\newline

\bibliographystyle{IEEEbib}
\bibliography{IEEEabrv,referencesAll}

\end{document}